\documentclass[preprint,aps,pre,nofootinbib]{revtex4-2}

\usepackage[utf8]{inputenc}
\usepackage{epsfig}
\usepackage{subfigure}
\usepackage{latexsym}
\usepackage{amsmath}
\usepackage{amsthm}
\usepackage{psfrag}
\usepackage{fullpage}
\usepackage{graphicx}
\usepackage{amssymb}
\usepackage{color}
\usepackage{placeins}
\usepackage{svg}
\usepackage{hyperref}
\usepackage{array}
\usepackage[autostyle]{csquotes}
\newcolumntype{x}[1]{>{\centering\hspace{0pt}}p{#1}}
\newcommand{\s}[1]{{\mathcal #1}}
\usepackage[normalem]{ulem}
\newcommand{\tsb}[1]{\textnormal{#1}}

\newcommand{\avg}[1]{\left\langle #1\right\rangle}

\newcommand{\T}[1]{\textrm{#1}}

\newcommand{\V}[1]{{\bf #1}}

\usepackage{float}
\theoremstyle{definition}

\theoremstyle{remark}

\usepackage{xr}
\makeatletter

\begin{document}
\title{Three-dimensional shape and connectivity of physical networks}
\author{Luka Blagojevi\'c}
\email{bl.luka@proton.me}
\affiliation{Department of Network and Data Science, Central European University, Vienna, Austria} 
\author{M\'arton P\'osfai}
\email{posfaim@ceu.edu}
\affiliation{Department of Network and Data Science, Central European University, Vienna, Austria}

\date{\today}

\begin{abstract}
Data describing the three-dimensional structure of physical networks is increasingly available, leading to a surge of interest in network science to explore the relationship between the shape and connectivity of physical networks. We contribute to this effort by standardizing and analyzing 15 data sets from different domains. Each network is made of tube-like objects bound together at junction points, which we treat as nodes, with the connections between them considered as links.
We divide these networks into three categories: lattice-like networks, trees, and linked trees. The degree distribution of these physical networks is bounded, with most nodes having degrees one or three.
Characterizing the physical properties of links, we show that links have an elongated shape and tend to follow a nearly straight trajectory, while a small fraction of links follow a winding path. These typical node and link properties must be reflected by physical network models. We also measure how confined a link is in space by comparing its trajectory to a randomized null model, showing that links that are central in the abstract network tend to be physically confined by their neighbors. The fact that the shape and connectivity of the physical networks are intertwined highlights that their three-dimensional layout must be taken into account to understand the evolution and function of physical networks.
\end{abstract}
\maketitle
\newpage

\section{Introduction}

Recently available detailed maps of physical networks provide an opportunity to systematically investigate the relationship between their shape and their network properties.
For example, open-data large-scale experimental projects provide a neuron-level mapping of biological neural networks~\cite{xu2020connectome,shapson2021connectomic}, 
high-throughput MRI measurements map out detailed vascular networks~\cite{gagnon2015quantifying},
or mycelia mapping projects explore local morphological and mechanical properties of fungi by representing them as networks of filaments~\cite{islam2017morphology}.
The nodes and links of these physical networks have two distinguishing features: (i)~they are characterized by a complex three-dimensional shape and (ii)~they physically interact with each other, for example, nodes and links obey volume exclusion, i.e., they cannot overlap.
To understand how physicality affects network evolution and function, we must extend the toolset of network science to take these features into account.

Recent work investigated artificial spatial embeddings of complex networks that obey volume exclusion~\cite{dehmamy2018structural}, the entanglement of physical links~\cite{liu2020isotopy}, models of physical network growth~\cite{posfai2024impact,pete2024physical}, and the effect of physical shape on the dynamics on networks~\cite{pete2024physical}.
However, systematic exploration of the three-dimensional shape, the network properties, and the relationship between them in real networks is still lacking.
Such exploration is hampered by the lack of standardized representation.
First, there is the technical difficulty that experimental maps of physical networks, like neural or molecular networks, are collected, processed, and analyzed with domain-specific methodology.
Therefore, any investigation of physical networks must be preceded by the time-consuming and computationally burdensome task of data pre-processing.
Second, even seemingly simple questions like what is a node and a link in a physical network carry a level of ambiguity: a physical network is a continuous object in space; to represent it as a network, we must discretize it into nodes and links.
The definition of nodes and links in turn affects, for example, what properties of the network we can study or the right choice of null models.

Here, we compile and standardize 15 data sets from various domains.
Each of these physical networks is composed of tube-like objects bound together at junction points; motivating us to treat the junction points as nodes and the tubes connecting them as physical links.
We characterize both the physical shape and the abstract network structure and the correlations between them.
For this, we calculate standard descriptors such as the degree distribution of the abstract network or the fractal dimension of the layout.
We also introduce a measure of link confinement to understand the role of volume exclusion, which compares the physical links to a null model that randomizes link trajectories.
The remainder of the paper is organized as follows:
In the next section, we describe the data sets we collected and their standardization.
In sections~\ref{network_section}-\ref{correlation_section}, we analyze the data sets' abstract network properties, their physical shape, and the emergent correlations between network and shape.
Finally, Sec.~\ref{discussion} provides a brief discussion.

\begin{figure}[h]
	\centering
	\includegraphics[width=1\textwidth]{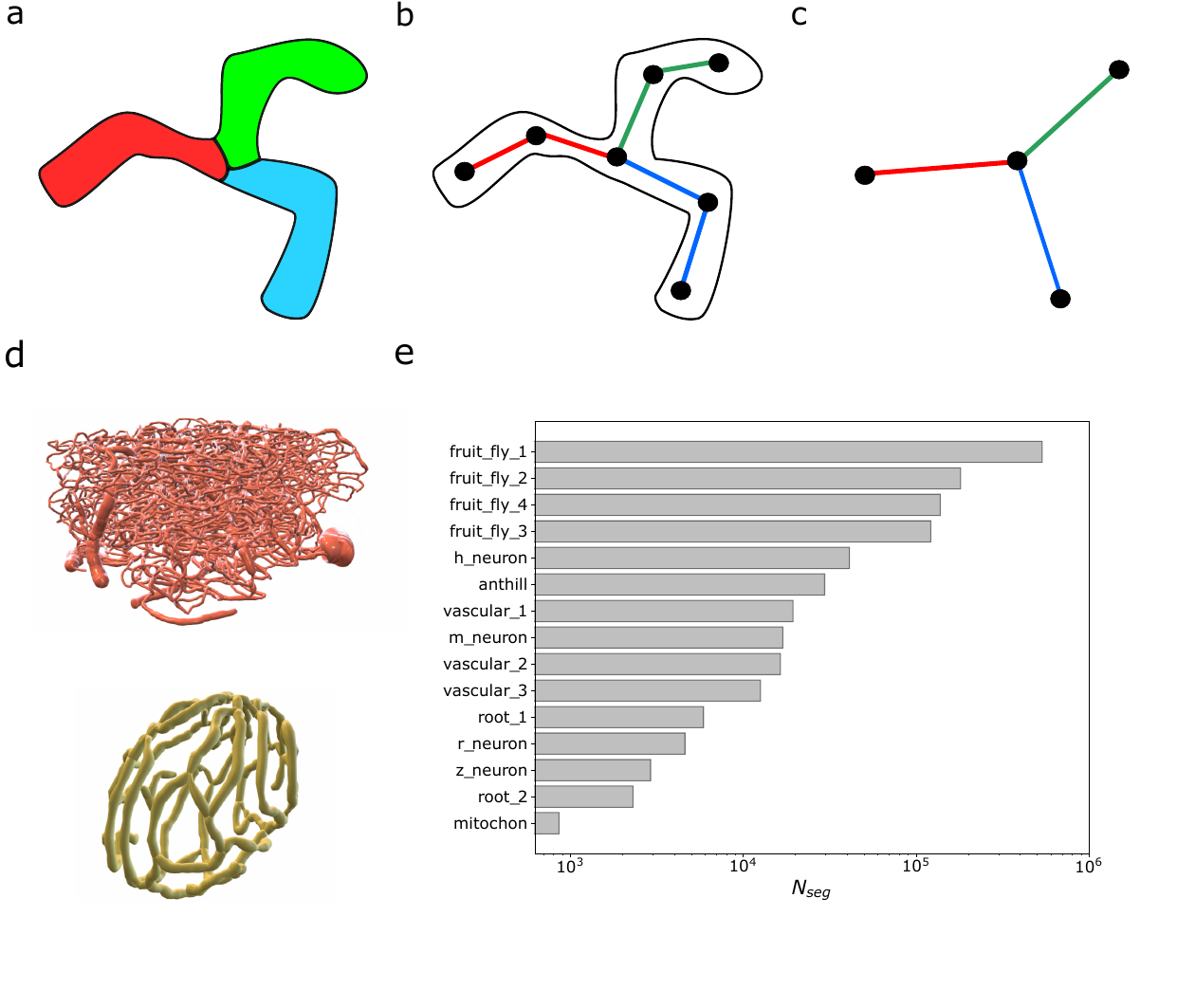}
	\caption{\textbf{Physical networks.}~\textbf{(a)} A physical network composed of three tube-like links bound together at a single junction point.
	\textbf{(b)}~The skeleton representation of (a) approximates the original structure as a collection of vertices (black points) and edges (colored segments).
	\textbf{(c)}~The combinatorial or abstract network of (a) captures the connectivity of the system without the physical structure: nodes represent junction points and terminal points, with a link between two nodes that are directly connected by a physical link.
	\textbf{(d)}~The skeleton representation allows us to approximate the original volume of real physical networks -- e.g., vascular network (top) and mitochondrial network (bottom).
	\textbf{(e)}~We compiled a set of 15 physical networks from various domains. The size of the networks varies greatly: the number of skeleton segments $N_{\tsb{seg}}$ capturing the shape of each network spans 4 orders of magnitude.}
	\label{definitions}
\end{figure}
\clearpage

\section{Data}\label{data_section}

Our goal is to systematically study the three-dimensional layout of physical networks and to understand the relationship between their physical properties and their network structure.
For this, we collected 15 data sets from various domains, including individual neurons~\cite{coskren2015functional,bartho2007cortical,koch2016big,torvund2017cone}, biological neural networks~\cite{scheffer2020connectome}, plant roots~\cite{ohashi2019reconstruction}, vascular networks~\cite{gagnon2015quantifying}, a mitochondrial network~\cite{viana2020mitochondrial}, and the imprint of an anthill (see SI S1).
Before any analysis, however, we must uniformly represent and standardize these data sets. In the following sections, we propose the use of a labeled skeleton representation, which efficiently captures both the physical shape and abstract network of physical networks.

\subsection{Skeleton representation}

Experimental imaging techniques that are used to capture the shape of physical networks, such as scanning electron microscopy or magnetic resonance imaging, typically output a three-dimensional image composed of voxels.
Hence, this voxel representation is the most accurate description available of network layouts. 
A three-dimensional voxel image, however, is difficult to handle both computationally and analytically; therefore a more compact representation of the data is needed.

Physical networks are typically composed of tube-like objects bound together at junction points (Fig.~\ref{definitions}a), making them suitable to be approximated by a series of straight segments and radii of the network at the endpoints of the segments (Fig.~\ref{definitions}b).
The process of creating these segments from a raw data format is called skeletonization~\cite{saha2016survey}, often employed in the fields of computer graphics~\cite{tagliasacchi20163d} and neuroscience~\cite{siddiqi2008medial,au2008skeleton}.  
Formally, a skeleton representation $\s S$ of a physical network is a graph whose set of vertices $\s V$ correspond to points in space and set of edges $\s E$ correspond to segments connecting point pairs.
Therefore each vertex $i\in \s V$ has a set of coordinates $\V r_i = (x_i,y_i,z_i)$ and a radius $\rho_i$ associated to it.
Figure~\ref{definitions}b shows the skeleton of Fig.~\ref{definitions}a.

The radius $\rho_i$ captures the thickness of the physical network at each vertex $i$.
Therefore, we can approximate the original occupied volume of a segment connecting vertex $i$ with radius $\rho_i$ and vertex $j$ with $\rho_j$ as a truncated cone, and the full volume of the network is approximated by the sum of these truncated cones. More specifically, for each edge $(i,j)$ in $\s S$ we add a truncated cone with axis $(\V r_i,\V r_j)$ and parallel faces with radii $\rho_i$ and $\rho_j$ which has volume:
\begin{equation}\label{eq:segment_volume}
V_{\tsb{seg}}(i,j) = \frac{1}{3} \cdot \pi \cdot (\rho_i^{2} + \rho_j^{2} + \rho_i\cdot \rho_j) \cdot |\V r_i -\V r_j|.
\end{equation}

Three-dimensional physical network data obtained from experiments is routinely skeletonized, and the skeleton of the network is published together with the raw data.
In fact, all but one of the 15 data sets that we study here was skeletonized by the original authors, the only exception is the anthill imprint provided to us as a surface mesh which we skeletonized using the Tangent-ball~\cite{whited2009tangent} algorithm from the \verb|Skeletor| Python module~\cite{au2008skeleton}.
The experimental setup and the choices made during the skeletonization may affect our analysis which is performed on the skeleton.
For example, increasing the number of skeleton segments $N_{\tsb{seg}}$ increases how well the skeleton approximates the original shape of the network.
However, increasing $N_{\tsb{seg}}$ also increases the size of the data set, hence increasing the computational burden of the analysis.
To improve the uniformity of the data sets, we perform two pre-processing steps:
\begin{enumerate} 
\item \textbf{Merging segments:} If two consecutive segments $(\V r_i,\V r_j)$ and $(\V r_j,\V r_k)$ are appear parallel to each other in the data set, we merge them into a single segment $(\V r_i,\V r_k)$ (see SI~S2.2).
\item \textbf{Skeleton healing:} Due to noisy data, a skeleton may be disconnected even when it represents a single continuous object in reality.
For example, a skeleton of a neuron may appear to have multiple components.
To remedy this, we add a segment to connect the two nearest skeleton vertices from two disconnected components. We repeat this step until the skeleton becomes connected. 
\end{enumerate}

Following these pre-processing steps, the number of segments $N_{\tsb{seg}}$ in a skeleton, as shown in Fig.~\ref{definitions}e, spans three orders of magnitude from mitochondrial networks, which have approximately~$10^3$ segments, to fruit fly neural networks which have up to~$10^6$ segments.

\subsection{Network structure}

A physical network is a continuous object embedded in Euclidean space, to characterize this object as a network we must separate it into discrete nodes and links.
For this, first note that all 15 data sets that we collected can be seen as a collection of tube-like objects bound together at junction points.
Hence, we define the junction and terminal points of the tubes as physical nodes and the non-branching tubes pairwise connecting these terminal and junction points as physical links.
A motivation for this definition is that cutting a physical link (i.e., a tube) at any point along its length causes the same disruption to the connectivity of the network.

More formally, for a skeleton representation $\s S$ of a physical system, we define each physical node to correspond to a skeleton vertex $i$ with degree $k(i)\neq 2$, and each physical link to corresponds to a path in the skeleton given by the ordered set $\s T(i_0,i_l)=[(i_0,i_1),(i_1,i_2),\ldots,(i_{l-1},i_l)]$, such that $k(i_0),k(i_l)\neq 2$ and $k(i_j)=2$ for $j=1,2,\ldots,l-1$.

With the above definition of nodes and links, we can talk about the abstract or combinatorial network $\s G$ of the system which captures its connectivity without the physical structure.
The skeleton $\s S$ is one possible physical realization of the abstract network $\s G$; however, there are many possible physical realizations of the same $\s G$.
In general, we are interested in understanding the relationship between the physical layout captured by $\s S$ and the network structure captured by $\s G$. 

As an example consider the physical network shown in Fig.~\ref{definitions}a which consists of three tubes bound together at a single junction point.
Its skeleton representation (Fig.~\ref{definitions}b), therefore, has three vertices with degree 1 corresponding to the terminal points, one vertex with degree $k=3$ corresponding to the junction point, and several vertices with degree $k=2$  tracing the trajectory of the tubes.
This means that the network consists of four physical nodes and three physical links, and its abstract network is a star (Fig.~\ref{definitions}c). 

Finally, note that for a given skeleton $\s S$, the above definition of physical nodes and links is not the only viable definition.
For example, in a neural network, it is natural to treat a neuron as a physical node and synapses between them as links, as individual neurons can have complex three-dimensional shapes that can be represented by a skeleton itself.
More generally, subgraphs of a large $\s S$ may represent functional units in a physical network and it can be useful to treat these functional units as physical nodes~\cite{pete2024physical}.
Note, however, that our definition of the abstract network $\s G$ provides the most detailed picture of the system's connectivity, and other definitions can be thought of as coarse-grained versions of $\s G$.

With the skeleton representation and the definition of the abstract network at hand, we are in the position to start our analysis.
In the following sections, we first explore the structure of the abstract networks of the 15 data sets, then we continue by characterizing their physical properties, and finally, we investigate the relation between the two.

\section{Abstract network properties}\label{network_section}

The above definition of physical nodes and links allows us to explore the properties of the abstract networks capturing the connectivity of physical networks without their three-dimensional structure; we focus on the degree distribution and motif frequencies.

\begin{figure}[h]
	\centering
	\includegraphics[width=1\textwidth]{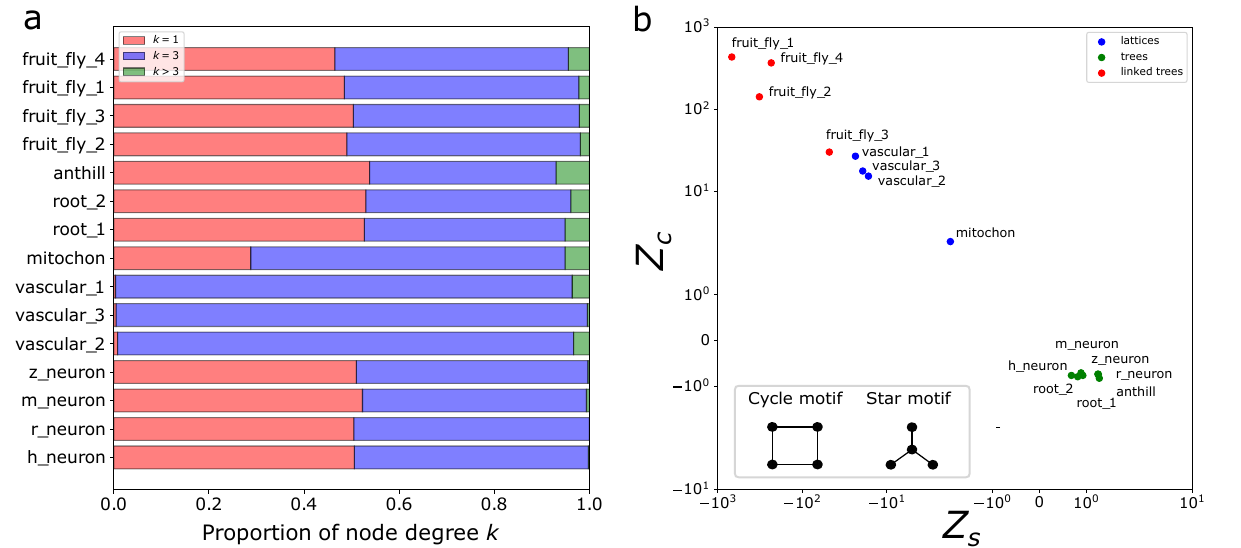}
	\caption{\textbf{Abstract network properties}~\textbf{(a)} The abstract networks are composed of terminal and bifurcation points; therefore, their degree distribution is mostly concentrated on $k=1$ and $k=3$.
	More specifically, lattice-like networks, such as vascular and mitochondrial networks, are mostly made up of branching nodes (degree $k=3$). For the rest of the network, nodes with $k=1$ and $k=3$ are approximately evenly split, as expected for tree networks.
	\textbf{(b)} We calculate the $z$-scores of four-node star and cycle motifs of the original networks compared to random networks with the same degree distribution. Markers are the average of $n_{\tsb{t}} = 200$ independent randomizations, and error bars representing the standard error of the mean are smaller than the marker size. Both axes are symlog axes, i.e., linear from -1 to 1, log otherwise for both positive and negative values.
	The star motif is slightly over- and the cycle motif is slightly under-represented in tree networks (green markers), as expected.
	In contrast, stars are under and cycles are overrepresented in both lattices (blue markers) and linked trees (red markers). The high $z$-scores observed for the 4-cycle indicate a lattice-like structure.}
 	\label{abstract network}
\end{figure}

The nodes in our physical networks are terminal and junction points, meaning that by construction nodes cannot have degree $k=2$, only degree $k=1$ or $k>2$.
Figure~\ref{abstract network}a shows the degree distribution $p(k)$ of all 15 networks, and our main observation is that most nodes have degree $k=1$ or $k=3$, and nodes with a larger degree are exceedingly rare, i.e., $p(1)+p(3)\approx 1$.
This means that junctions tend to be bifurcation points along the tube-like physical links making up the network.
This is in line with previous empirical observations and theoretical predictions for neurons~\cite{percheron1979quantitative, desai2022axon} and transport networks~\cite{labarbera1990principles,durand2006architecture}.
The observed narrow degree distribution is in contrast with degree-heterogeneous networks typically in the focus of network science and should be accounted for by mathematical models of physical networks~\cite{posfai2024impact}.

In graph topological terms, 7 out of 15 collected physical networks are trees: the individual neurons, the anthill imprint, and the plant roots.
Each bifurcation point with degree $k=3$ in a tree creates one new leaf node with degree $k=1$; therefore the fact that these networks are trees together with the observation that most junctions are bifurcation points, completely determines their degree distribution as $p(1)\approx p(3)\approx 1/2$.
The remainder of the networks contain cycles:
The vascular networks have no terminal points, apart from a few appearing due to finite sample size; therefore are almost completely composed of $k=3$ nodes.
The mitochondrial network representing a network of molecular strands has the highest fraction of $k>3$ nodes.
Finally, the fruit fly neural networks represent a collection of individual neurons which are trees, bound together by synapses, and their degree distribution closely resembles that of trees.

To explore the local loop structure of the networks, we calculate the abundance of observed 4-node motifs.
Namely, we focus on two motifs: the star motif and the 4-cycle.
To quantify their abundance, we calculate their Z-score compared to their degree-preserving randomized counterparts:
\begin{equation}
z_\T{s/c} = \frac{n_\T{s/c}-\avg{n_\T{s/c}}}{\sigma_\T{s/c}},
\end{equation} 
where $n_\T{s/c}$ is the number of occurrences of the star and 4-cycle motifs in the original networks, and the expected value $\avg{n_\T{s/c}}$ and standard deviation $\sigma_\T{s/c}$ is estimated by creating 200 independent randomizations~\cite{milo2002network}, while keeping the degree sequence fixed.
Figure~\ref{abstract network}b shows the scatter plot of $z_\T s$ and $z_\T c$ for the 15 networks.
As expected, in tree networks containing no cycles the star motif is slightly over-represented $z_\T{s}>0$, and the cycle motif is slightly under-represented $z_\T{s}<0$,
while for networks containing cycles, we find the opposite.
The highest abundance of 4-cycles is observed for the networks representing different brain regions of the fruit fly brain.

Based on the degree distributions and motif profile of the abstract networks, the physical networks fall into three broad categories: (i)~topological trees (the individual neurons, the root systems, and the anthill tunnel imprint), (ii)~lattice-like networks that are characterized by a loopy structure and few terminal points (the vascular networks and the mitochondrial network), and (iii)~linked trees which are a collection of trees bound together by additional links (the fruit fly brain regions).

\clearpage

\section{Physical properties}\label{physical_section}

In the previous section, we measured properties of the abstract network capturing the connectivity of the 15 physical networks.
We continue our investigation focusing on the physical properties of the system: we characterize the three-dimensional shape of the system without considering the abstract network.

\subsection{Space filling and fractal dimension}

\begin{figure}[h]
	\centering
	\includegraphics[width=1\textwidth]{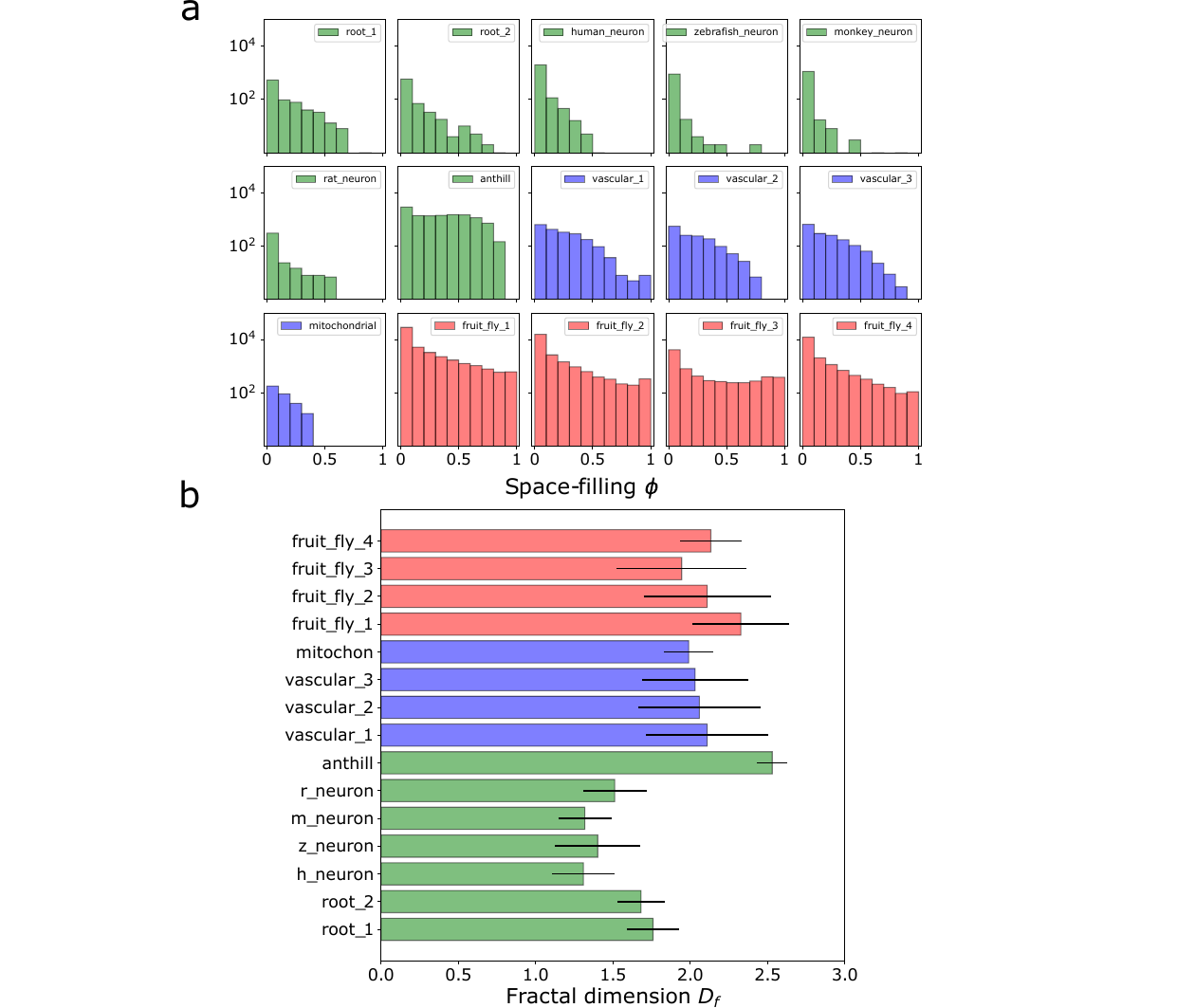}
	\caption{\textbf{Space-filling and fractal dimension.}
	\textbf{(a)} The boxplots of local space-filling $\phi$ show the coexistence of mostly sparse and fewer dense regions,  where the vertical lines indicate the median, the boxes span the $25$th-$75$th percentile range, and the whiskers extend to 1.5 times the interquartile range.
	\textbf{(b)} The fractal dimension of physical networks $D_\T{f}$ ranges between $1$ and the embedding dimension $D=3$. Linked trees (red) and lattices (blue) have fractal dimension values between $D_\T{f} \approx 2.0$ and $D_\T{f} \approx 2.3$, while trees (green) have more variation, ranging from $D_\T{f} \approx 1.5$ to $D_\T{f} \approx 2.0$ (except for from the anthill imprint). We estimate $D_\T{f}$ using the box-counting method, the error bars indicate the standard deviation of the local scaling estimates (see SI~S4.2).}
 	\label{network_physical_measures}
\end{figure}

A fundamental property of a physical network is its space-filling, i.e., the amount of volume it occupies from the available space.
We expect that networks that are tightly packed in space are strongly affected by physicality~\cite{dehmamy2018structural,posfai2023impact,pete2024physical}, although the study of random physical networks built from straight links suggests that volume exclusion can play a significant role even for diminishing small space filling~\cite{posfai2023impact}.
Many real physical networks have irregular shapes, hence much of their bounding box is unoccupied.
Therefore, instead of measuring space-filling globally, we divide the axis-aligned bounding box of each physical network into rectangular boxes.
We then measure the local space-filling in each box $i$ as
\begin{equation}
\phi(i) =\frac{V_\T{occ}(i)}{V_\T{box}},
\end{equation}
where $V_\T{occ}(i)$ is the volume of the intersection of the network and box $i$ and $V_\T{box}$ is the volume of the rectangular box.
The distribution of $\phi(i)$ depends on the choice of $V_\T{box}$; therefore, to ensure consistency, we set them separately for each data set such that every bounding box is split into a $10\times10\times10$ grid of boxes (see SI~S4.1.).
Figure~\ref{network_physical_measures}a shows the distribution of $\phi$ for all 15 physical networks, revealing that the physical networks fill out the space with mostly sparse regions and fewer denser regions.
Therefore we expect that physicality will also affect the network structure unevenly: volume exclusion can limit the number and shape of links in dense regions. 

To further characterize the shape of the networks, we calculate their box-counting fractal dimension $D_{\tsb{f}}$, which compactly describes the space-filling of a physical object on multiple scales~\cite{soddell1994using,foroutan1999advances,vicsek1992fractal} and is widely used to characterize the shape of complex biological systems~\cite{vicsek1992fractal, bunde2013fractals}.
Possible values of $D_{\tsb{f}}$ for connected networks range between 1 and the embedding dimension $D = 3$,
Fig.~\ref{network_physical_measures}b shows $D_{\tsb{f}}$ for the 15 networks, we find that both lattices and linked trees have $D_{\tsb{f}}\gtrapprox 2$, while trees are typically characterized by $D_{\tsb{f}}\lessapprox 2$, except for the anthill imprint.
The fractal scaling spans at least two orders of magnitude of length scales (see SI~S4.2.), again pointing towards regions of high and low physical density at different resolution levels.

\clearpage

\subsection{Link volume and shape}
\label{volume_shape_sec}
The distribution of space-filling and the fractal dimension characterize the shape of physical networks as a whole.
In this section, we continue by quantifying the shape of individual links; we focus on their volume and their straightness.

\begin{figure}[h]
	\centering
	\includegraphics[width=.8\textwidth]{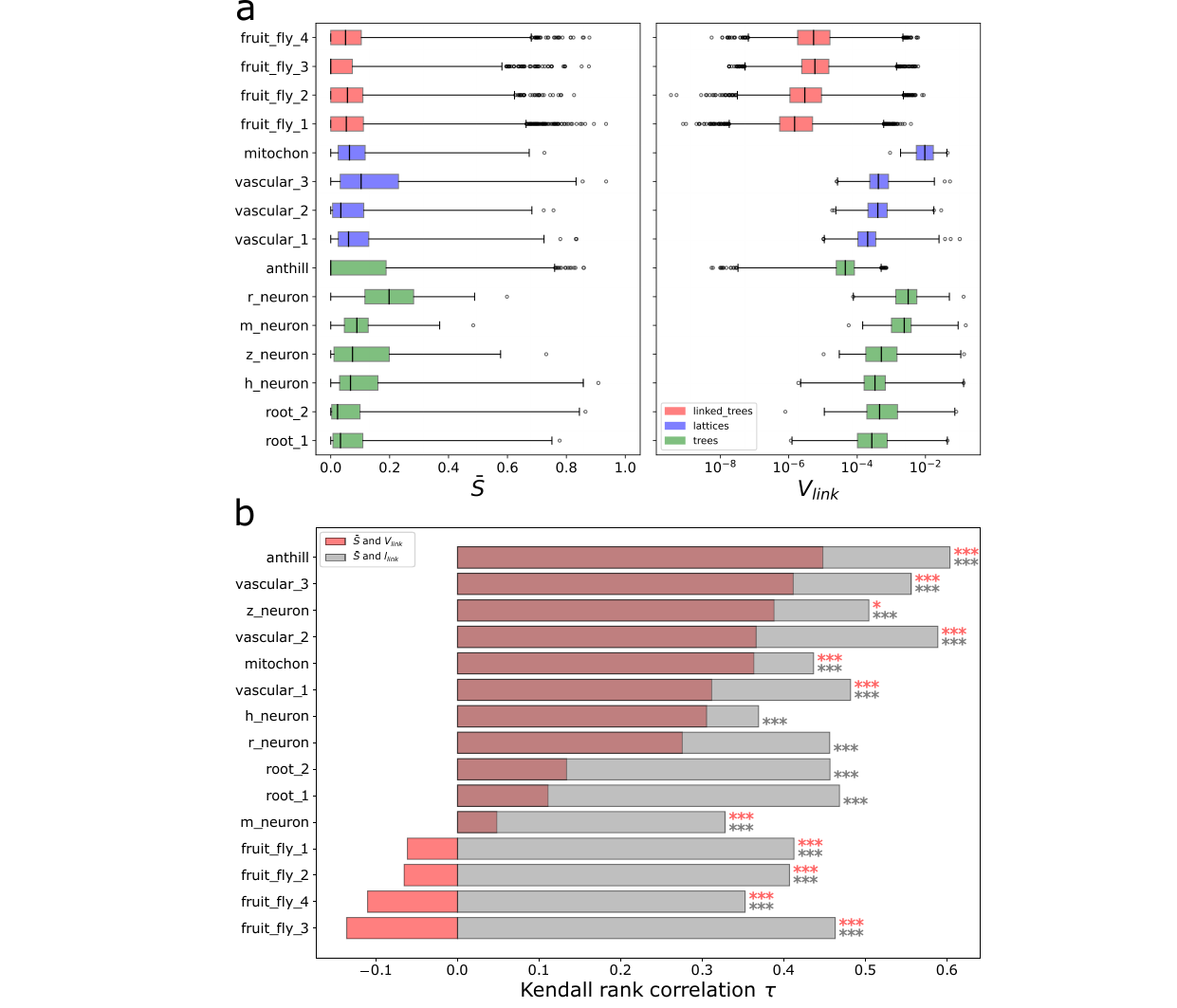}
	\caption{\textbf{Link shape, length and volume.} 
	\textbf{(a)} Box plots of $\bar{S}$ and $V_{\T{link}}$, where the vertical lines indicate the median, the boxes span the $25$th-$75$th percentile range, and the whiskers extend from the $0.1$th to the $99.9$th percentile.
	We find that the bulk of the distributions fall in the range between $\T{med}(\bar{S}) =0$ and $ \T{med}(\bar{S})= 0.1$, indicating that the networks are mostly composed of close-to-straight links.
 Link volume $V_{\T{link}}$  distributions span a wider range for linked trees and are the most narrow for lattices.
	\textbf{(b)} By computing Kendall rank correlation $\tau$ between link complementary straightness $\bar{S}$, link volume $V_{\T{link}}$ and link trajectory length $l_{\T{link}}$, we observe a consistent trend of $\tau > 0$, meaning that longer and more voluminous links tend to have more winding paths.
	This trend is only reversed for the fruit fly networks, which have $\tau < 0$ between $\bar{S}$ and $V_{\T{link}}$. This could be explained by neuron somatas, which are represented as high-volume, straight links composed of a small number of skeleton segments. }
 	\label{link_shape_volume}
\end{figure}

The skeleton describing the three-dimensional shape of a physical network is composed of straight segments connecting pairs of skeleton vertices and a radius associated to each skeleton vertex, allowing us to approximate the volume belonging to a segment in Eq.~\eqref{eq:segment_volume} as a truncated cone.
The total volume of a link $(i,j)$ is then:
\begin{equation}
V_\T{link}(i,j) =\frac{1}{V_\T{total}}\sum_{(v,w) \in (i,j)}{V_{\tsb{seg}}(v,w)},
\end{equation}
where $V_{\tsb{seg}}(v,w)$ is the volume of each segment tracing the link $(i,j)$. We normalize the link volume by the total volume of the network $V_\T{total}$, setting the unit of measurement.
We find that linked trees, or the fruit fly neural networks, have consistently high link volume heterogeneity, as their distributions span 6 to 8 orders of magnitude, which is higher compared to most lattices and trees (Fig.\ref{link_shape_volume}a).

We also measure the aspect ratio of physical links $a(i,j)=\rho_\T{link}(i,j)/l_\T{link}(i,j)$, where $\rho_\T{link}$ is the average radius and $l_\T{link}(i,j)$ is the length of link $(i,j)$.
We find the largest average aspect ratios for one of the fruit fly networks and the anthill ($\text{med}(a) \approx 0.3$), while $\text{med}(a)$ is substantially lower for other data sets (see SI~S1.1).
Overall, this confirms that physical links are elongated tube-like objects.

Since physical links are tube-like objects, we can capture most of their shape by characterizing their one-dimensional trajectory.
Here, we calculate the deviation of the link trajectories from a straight line, quantifying how curved a link is.
For this, we rely on a measure of straightness introduced originally in the context of geographical networks~\cite{crucitti2006centrality}, namely we calculate the complimentary straightness for each link
\begin{equation} 
\bar{S}(i,j) = 1 - \frac{\lvert\V r_i - \V r_j\rvert}{l_{\tsb{link}}(i,j)},
\end{equation}
where $\lvert\V r_i - \V r_j\rvert$ is the Euclidean distance between nodes $i$ and $j$ and $l_{\tsb{link}}(i,j)=\sum_{(v,w) \in (i,j)}\lvert\V r_v - \V r_w\rvert$ is the length of the physical link $(i,j)$.
The complimentary straightness $\bar{S}(i,j)$ is 0 if the physical link is straight and close to 1 if it follows a winding trajectory much longer than the straight path between the two points.

Calculating the median of complimentary link straightness distribution, $\text{med}(\bar{S})$, reveals that links in all of the 15 physical networks tend to follow a trajectory close to a straight line: most networks have $\text{med}(\bar{S})\approx 0.1$, meaning that the length of links is most often less than 10\% longer than the optimal straight trajectory.
Similarly to link volume heterogeneity, linked trees tend to cluster together and are among the networks with the straightest links with $\text{med}(\bar{S}) = 0.05$.
Reference~\cite{posfai2023impact} introduced random linear physical networks, a minimal model that constructs a physical network from straight cylinders.
The fact that we observed an abundance of straight or close-to-straight links lends support for using such linear physical network models to understand the role of physicality in real networks. Note, however, that although most links are close to straight, the distribution of $\bar{S}$ is right-skewed as seen in Fig.~\ref{link_shape_volume}, which points to a smaller fraction of links that significantly deviate from a straight trajectory (see SI~S3.).

Finally, we computed the correlations between link straightness $S(i,j)$ and the total link length $l_\T{link}(i,j)$ and volume $V_\T{link}(i,j)$ for each data set using Kendall's rank correlation coefficient $\tau$~\cite{abdi2007kendall}.
Figure~\ref{link_shape_volume}b shows that for all networks, there is a positive rank correlation $\tau>0$ between $\bar{S}(i,j)$ and $l_\T{link}(i,j)$, indicating that longer links tend to follow a more winding path.
We also observe a positive correlation $\tau>0$ between $\bar{S}(i,j)$ and $V_\T{link}(i,j)$, since longer links tend to have larger volume.
The only exception to this is the fruit fly neural networks, for which $\bar{S}(i,j)$ and $V_\T{link}(i,j)$ are negatively correlated $\tau<0$.

One possible cause contributing to the negative correlation is that fruit fly neural networks are composed of neurons that have large somata, which are represented in the data set as short, yet high-volume segments (see SI~S1.7.).

\clearpage

\subsection{Link confinement}

In the previous sections, we found that space-filling and link properties are heterogeneously distributed: most regions of space are sparse and most links are close-o-straight, yet there exists dense regions of the network and a small fraction of links follow paths that deviate from a straight line significantly.
This suggests that volume exclusion or other repulsive physical interactions may also play an uneven role in shaping the network.
To further investigate this hypothesis we devise a quantity that captures the confinement of a link by other components of a network.

A link $(i,j)$ following a trajectory $\s T(i,j)$ in a real physical network obeys volume exclusion: it does not overlap with other links.
Our strategy to quantify the role of repulsive forces that may shape $\s T(i,j)$ is to calculate the number of overlaps with other links for a random ensemble of synthetic links that follow similar trajectories to  $\s T(i,j)$.
If the synthetic links typically overlap with many other links, the trajectory $\s T(i,j)$ is an outlier and must be shaped by forces not captured by the random ensemble.

The trajectory of a physical link $(i,j)$ in our data sets is given by the ordered set $\s T(i,j)$ of oriented three-dimensional segments.
To generate the random trajectory $\s T_\T r(i,j)$, we shuffle the order of the segments while maintaining their orientation and length, creating a uniform random permutation of $\s T(i,j)$.
The randomization preserves the endpoints and the total length of the link, but otherwise removes any correlation between subsequent segments (see SI~S4.4.); therefore, the possible link trajectories $\s T_\T r(i,j)$ have the same complimentary straightness $\bar{S}$ as the original link.
Next, we estimate $I(i,j;l,k)$, the expected number of intersections between the randomized link $\s T_\T r(i,j)$ and a non-randomized link $\s T(k,l)$.
To quantify the confinement of the link $(i,j)$, we sum up the expected number of intersections with other links:
\begin{equation}\label{eq:link_conf}
C(i,j) = \sum_{k,l\neq i,j}I(i,j;l,k)+I(l,k;i,j),
\end{equation}
where the first term corresponds to intersections when link $(i,j)$ is randomized, and the second term corresponds to intersections when link $(l,k)$ is randomized.
Note that the summation in Eq.~\eqref{eq:link_conf} excludes links that share an endpoint with $(i,j)$.
We do this to exclude trivial intersections from the count, since adjacent links $(i,j)$ and $(j,k)$ necessarily overlap at the junction point $j$ even for non-randomized link trajectories.

\begin{figure}[h]
	\centering
	\includegraphics[width=1\textwidth]{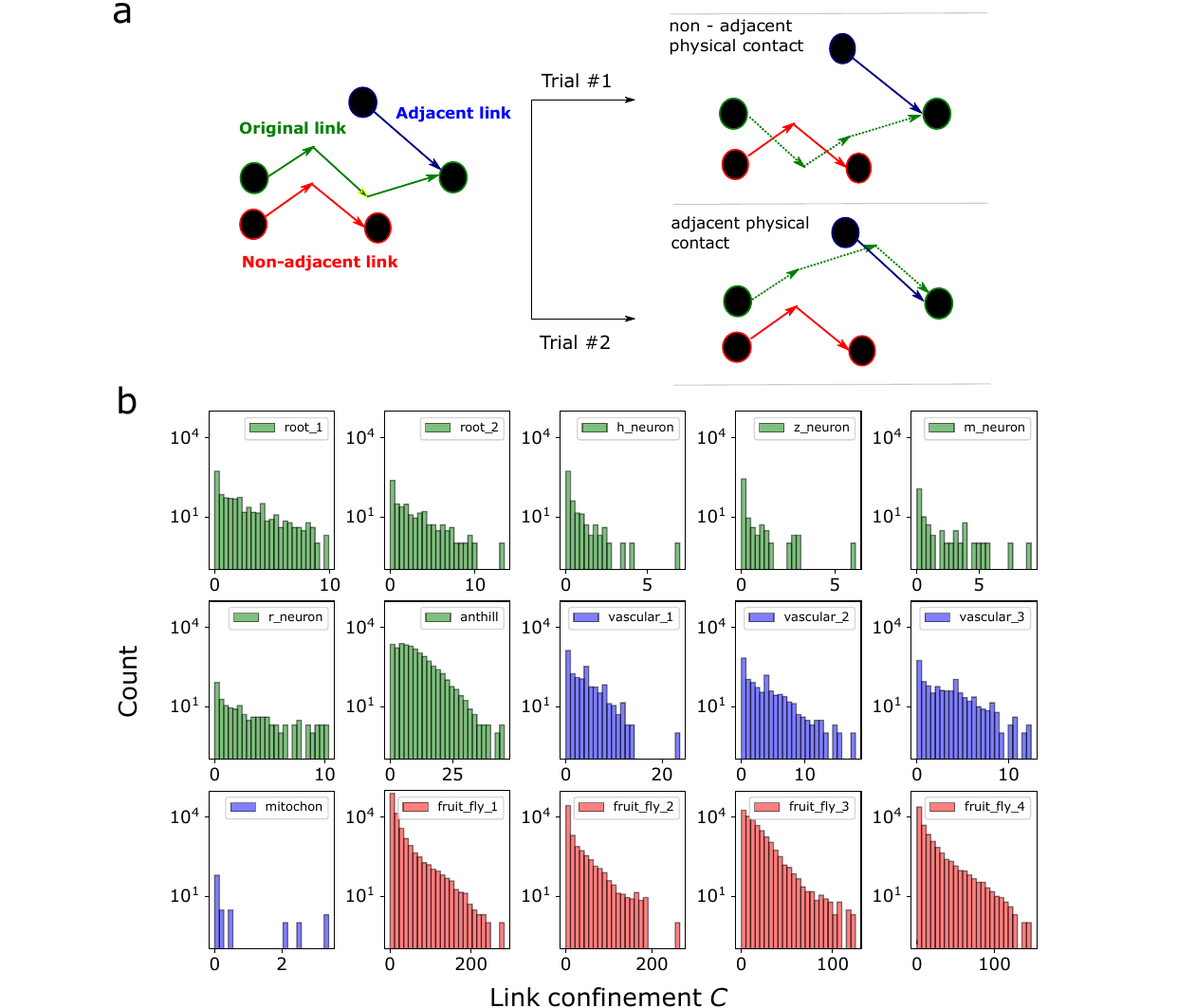}
	\caption{\textbf{Link confinement.}
\textbf{(a)} To quantify volume exclusion at the link level, we introduce link confinement $C$, the expected number of intersections after randomizing a link's trajectory.
To estimate $C$ of the green link, we randomize its trajectory twice. In trial \#1, the green link intersects the red link, while in trial \#2, it intersects the blue link. The blue link, however, shares an endpoint with the green link; therefore their intersection is not counted.
\textbf{(b)} The distribution of $C$ for each physical network.
There are many links with link confinement values close or equal to $C=0$, and typical values of link confinement are around $C\sim 10$, and only for the fruit fly networks, there are links with link confinement values $C > 100$, indicating highly confined links for the linked tree networks.}
 	\label{link_confinement}
\end{figure}

The procedure of calculating $C(i,j)$ is illustrated by Fig.~\ref{link_confinement}a: we start with a link (green) surrounded by two neighboring links (red and blue).
The figure shows $n_{\tsb{t}} = 2$ randomization trials of the green link:
In trial one, the randomized link intersects the red link, but not the blue.
In trial two, the randomization creates an intersection with the blue link; the two links, however, are adjacent (they share a junction point), hence the intersection is not counted.
In this particular example, the contribution to the link confinement measure from the randomization of the green link will be:
\begin{align}
I(\T{green},\T{red})&= \frac{0+1}{2} = 0.5\\
I(\T{green},\T{blue})&= \frac{0+0}{2} = 0
\end{align}
To complete the calculation of $C(\T{green})$, we also need to randomize the red link to estimate $I(\T{red},\T{green})$ in the same manner, while $I(\T{blue},\T{green})=0$ by definition.
Finally, the confinement of the green link is obtained by summing up the contributions, i.e., $C(\T{green})=I(\T{green},\T{red})+I(\T{red},\T{green})$.

Collision detection between link trajectories is a computationally expensive task, in practice we randomize each link $n_\T t=20$ times (and $n_\T t= 5$ for the fruit\_fly\_1 network) and we rely on an efficient collision detection algorithm leveraging kd-trees~\cite{schauer2015collision} (see SI~S4.4 and SI~S4.5).

Figure~\ref{link_confinement}b shows a large variation in link confinement $C$.
Across all networks, physical links are characterized by $C\approx 0$, indicating that these links are not affected by the physical proximity of other components of the network.
However, we also find highly confined links with $C > 10$ and even $C > 100$ expected intersections, suggesting again that physicality tends to play a heterogeneous role in forming networks.
In particular, the linked tree networks typically have heavy-tailed link confinement distributions (see SI~S5).
In terms of absolute counts, the linked trees or the fruit fly neural networks show the highest values of link confinement.
This can be explained by the fact that these networks are composed of multiple neurons, hence in these networks, we have more complete information about the physical environment of the links, compared to networks that describe single neurons.

\clearpage

\section{Link confinement correlation profiles}\label{correlation_section}

In the previous section, we defined the link confinement $C$ as the expected number of intersections if a link would follow a random trajectory, allowing us to identify links whose trajectory is most affected by repulsive forces in the network.
Here, we characterize the properties of such confined links by calculating the Kendall rank correlation $\tau$ between the link confinement $C$ and other link properties.
Specifically, we focus on the (i)~physical properties, complementary straightness $\bar{S}$ and link volume $V_\T{link}$ (ii)~abstract network properties link betweenness $B_\tsb{link}$  and link degree $k_{\tsb{link}}(i,j)$, where the latter is defined as the sum of the degrees of the endpoints of link $(i,j)$.

\begin{figure}[!htbp]
	\centering
	\includegraphics[width=1\textwidth]{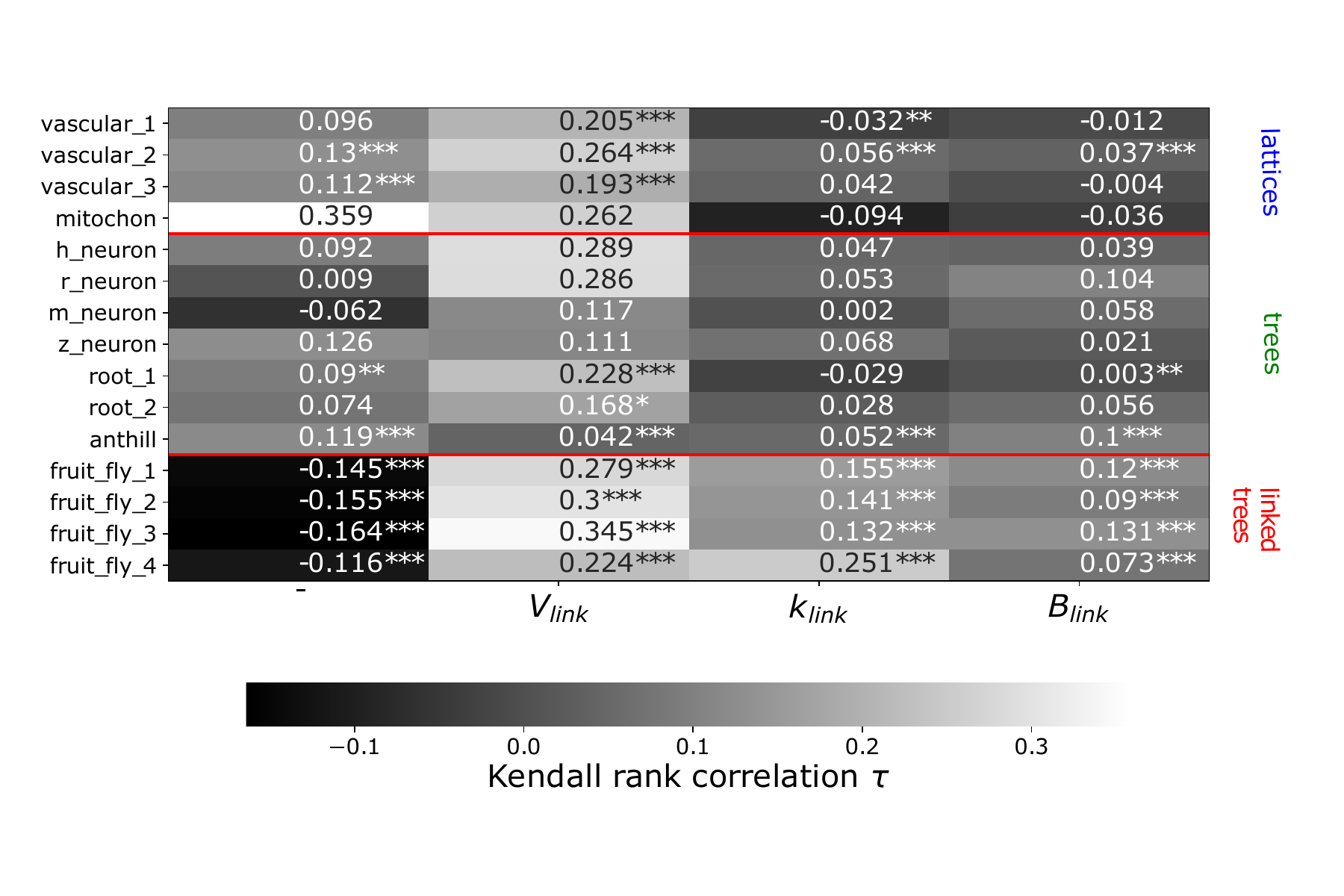}
	\caption{\textbf{Link confinement correlations.} We compute the Kendall rank correlation $\tau$ between the link confinement $C$ and the physical and the abstract network properties of links, obtaining a correlation profile for each physical network. Across all datasets, except some trees, link volume $V_{\T{link}}$ and link confinement $C$ have a statistically significant ($p<0.01$) moderate positive relationship. We also find a significant association with complementary straightness $\bar S$, which is positive for lattices and trees, and negative for linked trees, consistent with the correlations between $V_\T{link}$ and $\bar S$ (Fig.~\ref{link_shape_volume}).
 For the linked trees, which are the data sets with the most complete information about the environment of the physical links, we find a significant positive association between the centrality of the links in the abstract network (link degree $k_{\T{link}}$ and link betweenness $B_{\T{link}}$) and their link confinement $C$.}
	\label{correlation_heatmaps}
\end{figure}

Figure~\ref{correlation_heatmaps} shows the correlation profiles of all 15 networks.
A persistent pattern we observe is the positive correlation between link confinement $C$ and link volume $V_\T{link}$.
This is expected since larger links have more opportunities to intersect or be intersected by neighboring links. 
On the other hand, correlations between link confinement and straightness show a more curious pattern: we observe that most networks tend to have positive and significant correlations $\tau$ between the link confinement $C$ and link complementary straightness $\bar{S}$, as expected, indicating that more winding links are also more confined. However, for linked trees (fruit fly neural networks) we find a negative $\tau$ between $C$ and $\bar{S}$.
To explain this recall that in the fruit fly neural networks we found a negative correlation between link volume and link length and a positive correlation between straightness $\bar{S}$ (Sec.~\ref{volume_shape_sec}).
This means that short links tend to be more confined due to their large volume, while also follow a straighter path.

For correlations between link confinement $C$ and the abstract network properties, such as link betweenness $B_\tsb{link}$ and degree $k_\tsb{link}$, we find consistent and significant positive correlations for fruit fly neural networks.
This indicates emergent correlations between the three-dimensional layout and abstract network properties of physical networks: more central links in the abstract network tend to be more confined in physical space.
For lattice-like networks and trees, we find less consistent and weaker positive correlations.
Overall, we are able to show that the abstract network structure and physical layout are intertwined for networks where we have sufficient information about the surrounding environment of the physical links, such as the fruit fly neural networks, which contain multiple neurons in close proximity with each other.

\clearpage

\section{Discussion}\label{discussion}

Experimental data describing the three-dimensional shape of physical networks is increasingly becoming available, and the growth in the number and size of these data sets is expected to continue: connectome of the human brain consists of $\approx 10^9$ neurons and fungal mycorrhizal networks are estimated to span $\approx 10^{17}$ km in Earth’s soil~\cite{leake2017mycorrhizal}. 
The new data calls for extending the toolset of network science to analyze, model, and understand how the three-dimensional layout and physical interactions shape the structure and function of physical networks.
Here, we contributed to this effort in three distinct ways: (i)~We collected and standardized 15 data sets describing the three-dimensional layout of physical networks from diverse domains.
(ii)~We characterized the structure of both the abstract network and physical layout of the 15 systems using descriptors such as the degree distribution and fractal dimension.
(iii)~We introduced link confinement as a method to quantify how physical interactions shape link trajectories in physical networks, allowing us to investigate emergent correlations between physical and abstract network properties.

Our work may support future research on physical networks in several ways.
First, we promote the use of labeled skeleton graphs to represent both the layout and the connectivity of physical networks.
The skeleton captures the shape of the network, while the labeling identifies the physical objects corresponding the the nodes and links of the abstract network.
Here, we focused on treating junction points in the skeleton as nodes and sequences of segments connecting them as links; however, the labeled skeletons are not limited to such interpretation.
For example, sub-graphs representing larger functional units, such as neurons in the brain, can be identified as physical nodes.

Second, our results also inform theoretical models of physical network growth.
Recent work that modeled physical nodes as spheres and links between them as tubes~\cite{dehmamy2018structural,posfai2024impact}.
We found that most physical links follow close to straight trajectories, suggesting that linear physical network models where links are straight cylinders are indeed a useful class of models to understand physicality in networks.
On the other hand, these physical network models generalize the classic Erd\H{o}s-R\'enyi and Barab\'asi-Albert models to physical space and thus do not restrict the node degree.
We, however, found that junction points in real physical networks almost exclusively have degree three, a fact that must be accounted for by future models.
Note that to obtain real physical networks with non-trivial degree distributions one must abandon identifying junction points as nodes, instead we must identify larger sub-graphs of the skeleton as physical nodes.
In more formal terms, these networks can be modeled as a network-of-networks: we represent each physical node as a skeleton that has junction points with degree 3, and these physical sub-networks are bound together to form a network-of-networks with no restriction on the number of connections a sub-network can make with other sub-networks~\cite{pete2024physical}. Future work may explore the relationship between the network-of-networks representation and the more fine-grained junction network representation.

Finally, we quantified the physical confinement of individual links by comparing the path that links follow to randomized trajectories, allowing us to identify correlations between physical and abstract network properties. In general, understanding the relationship between physical shape and abstract network structure is one of the key challenges of physical network research~\cite{posfai2024impact,pete2024physical, flores2023network}. Future work may rely on other spatially randomized null models and abstract network measures to probe the relation between the two. 

Our work is limited by the scope of the available data sets and computational constraints.
First, our data sets do not contain information about the environment the networks are embedded in; therefore, we can only investigate interactions between the components of the network and not interactions between the networks and their surroundings.
For example, we found the strongest relationship between link confinement and abstract network structure for the fruit fly neural network data sets and we found a weaker or no relationship for individual neurons.
This is likely due to the fact that the fruit fly data sets contain multiple neurons, thus capturing more of the environment of individual physical links.
Future work may consider more complete data sets as they become available or theoretical models of network growth could incorporate non-trivial environments.

\subsection*{Data availability}
All data used in the manuscript is publicly available. Code to process the data is provided at: \href{https://github.com/lukablagoje/three-dimensional-shape-connectivity-physical-networks}{https://github.com/lukablagoje/three-dimensional-shape-connectivity-physical-networks}

\bibliographystyle{unsrt}
\bibliography{references}


\end{document}